\documentclass{article}

\usepackage{myllncs}
\usepackage{a4wide}
\usepackage{graphicx}
\usepackage{amsbsy}
\usepackage{amssymb}
\usepackage{subfigure}

\def\beq{\begin{equation}}
\def\eeq{\end{equation}}
\def\d{{\rm d}}
\def\balpha{\bar{\alpha}}
\def\smalS{{\scriptscriptstyle{(S)}}}
\def\sd{\strut\displaystyle}
\def\wmax{\overline{w}}

\begin{document}
\bibliographystyle{plain}

\title{Numerical analysis of frazil ice formation in
turbulent convection}
\author{A. Abb\`a\inst{1} \and M. Montini\inst{1}
\and L. Pignagnoli\inst{2,3} \and L. Valdettaro\inst{1} \and P. Olla \inst{4}}
%
%
\institute{Dipartimento di Matematica, Politecnico di Milano,
Piazza L. da Vinci 32,
I-20133 Milano, Italy
\and
ISAC-CNR, I-40129 Bologna, Italy
\and
Dipartimento di Matematica, Universit\`a degli Studi
di Milano, I-20133 Milano, Italy
\and
ISAC-CNR and INFN, Sez. Cagliari, I-09042 Monserrato, Italy
}

\date{\today}
\maketitle


\section{Abstract}
We study 
the first steps of ice formation in 
fresh water turbulent convection
(frazil ice regime).
We explore the sensitivity of the
ice formation process to the set of non-dimensional parameters
governing the system.
We model the mixture of ice crystals and water
as a two-phase medium composed of water and ice particles of fixed
diameter. 
We use the Boussinesq approximation and we
integrate numerically the set of equations
making use of a numerical code
based on second order finite difference.
A dynamic LES model for the subgrid scales is used.

We show that the ice particle
rise velocity and the ice concentration source term coefficient 
significantly affect the frazil ice dynamics.
The maximum of ice production is obtained
in those situations where the rise velocity is
of the same order of magnitude of a characteristic velocity
of the thermal downwelling plumes.
We develop a simple model which
captures the trend of the growth rate as a function
of the relevant parameters.
Finally we explore the nonlinear regime and we 
show that the parameter
which plays a key role in fixing the concentration
value at the statistically steady state is the
heat exchange source term coefficient.

\section{Introduction}\label{intro}

The first stage of ice formation at the supercooled free
surface of oceans, rivers and lakes 
is called frazil ice, a suspension of individual, small
randomly-oriented crystals typically measuring approximatively $ 1
\div 4 $ mm in diameter and $ 1 \div 100$ $\mu$m in thickness
\cite{Kivisild70},\cite{Martin81a}.

\begin{figure}[hbtp]
\begin{center}
\includegraphics[width=.7\textwidth]{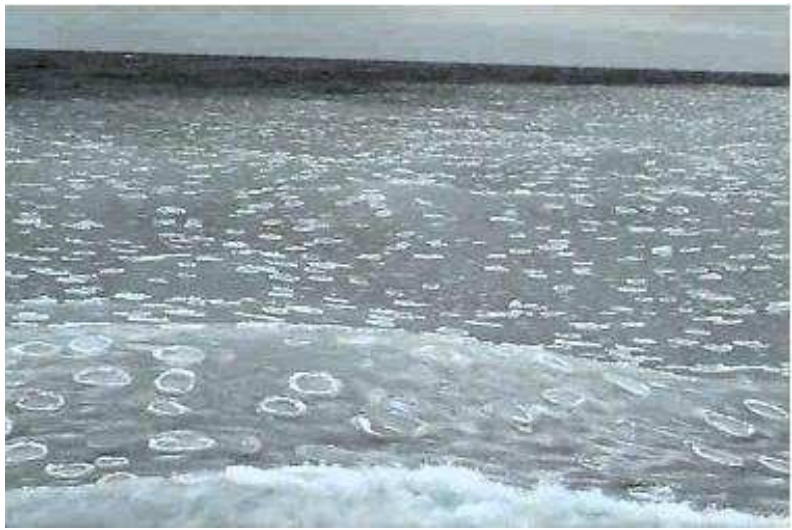}
\caption{Frazil-pancake ice field in Polar ocean
(www.vims.edu/bio/microbial/NBPice.html).}\label{pancake}
\end{center}
\end{figure}

Frazil ice evolves rapidly into a thin slurry of ice
platelets giving the sea a greasy or milky appearance, known as
grease ice. As
cooling continues and the density of the frazil suspension
increases, the frazil crystals freeze together into small cakes.
Collisions between neighbouring cakes forces frazil ice over the
cake edges where rims form to give the characteristic form of
pancake, hence the appellation pancake ice. Initially formed
pancakes may be only a few centimeters in diameter, but they grow
in diameter and thickness from surrounding frazil matrix and may
reach $3 \div 5 $ m in diameter and $ 50 \div 70$ cm in thickness.
The last stage of sea ice evolution is the coalescence of pancake
ice floes, which aggregate into a continuous ice sheet. In figure
(\ref{pancake}) a picture of a frazil-pancake ice field in Polar
ocean is reported.

{In rivers, frazil ice production is a transient
phenomena which occurs in turbulent supercooled water. The first
stage is the water temperature lowering to about $0.01 \div 0.1
\hspace{0.5mm}^{\circ}C$ below the freezing point because of the
wind action or the radiative cooling; then the frazil discs start
to form and the water temperature following the mass of crystals
increases in the order of minutes toward the freezing point. The
presence of frazil ice can cause serious problems to hydroelectric
facilities such as the blocking of turbine intakes, the blockage
of hydroelectric reservoirs and the freezing open of gates.}

In polar ocean, frazil ice can form under ice covers and
in open region within polar pack ice, more precisely in $100$
meters scale leads and in kilometer scale polynyas. Frazil ice is
typically present in the Marginal Ice Zone (MIZ), which is the
transition region between the open polar ocean and the continuous
ice that covers the central basin. {Due to the combined
action of wind and surface waves, the turbulent waters of MIZ
prevent the formation of large ice sheets. The presence of frazil ice
fields is associated with important geophysical and biological
processes of the polar oceans so that the study of frazil ice
dynamics is an active research field. As frazil ice production is
accompanied by salt rejection, it is believed that it could play
an important role in stimulating the convection process of ocean
waters. Moreover, frazil ice plays an important role as scouring
and sediment transfer agent and also it can transport biological
species from the sea bottom to the surface pack ice.}

This paper is organized as follows. Section \ref{Frazil}
is devoted to a brief description of the frazil ice formation
process. In section \ref{Model} the derivation of the frazil ice
transport model is reported, while section \ref{Boussinesq} is
dedicated to the Boussinesq-like approximation. In section
\ref{Results} we present numerical results of the simulations and
finally section \ref{Conclusions} is devoted to the conclusions.

\section{The formation of frazil ice}\label{Frazil}

The time it takes for ice formation in oceans, rivers
and lakes can be quite different from one year to the other, due
to meteorological variations. Estimating this time is a very
complicated task, which involves many parameter such as wind,
water temperature and water depth. Immediate applications of this
problem can be found in shipping and ice breaking service. To this
end, many theoretical and experimental studies were performed
during the last decades, giving useful information about time
histories of temperature and ice formation.

Designers of hydraulics structures such as canals, water
transport facilities and hydroelectric power structures are
particularly interested in three ice-flow regimes, which are
common in rivers during freeze-up period: continuous ice sheet,
floating frazil ice and well mixed flow (with frazil ice crystals
vertically mixed in the water flow). In the last case, if the
velocity of the river is sufficiently large, open water conditions
may persist throughout the winter. In consequence of this, a very
important task is to predict, starting from a given set of flow
conditions, which kind of regime could be expected, and the
transition from an ice cover to open water. Obviously, when frazil
ice crystals remain well mixed with the fluid, they cannot
accumulate beneath the surface and the ice cover can not form, so
the conditions under which the transition from well mixed flow to
floating ice layer occurs, could be employed as a rough criteria
for the formation of the continuous ice cover. {In
\cite{Gosnik83} the rough criteria cited above was compared, with
good agreement, with experimental and field data. In particular, it
has been remarked that the presence or absence of the well mixed
regime could be forecasted comparing the buoyancy time scale $T_B$
with the vertical turbulent diffusive mixing scale $T_D$. When
$T_B < T_D$ a layered flow will develop, because the frazil ice
crystals rise to the water surface faster than they are removed by
turbulent mixing. Under the same argument, the opposite case $T_B >
T_D$ corresponds to a well mixed flow.} More sophisticated
criteria take into account other hydrological aspects, such as
sinuosity, channel and slope variability, the transition from
frazil ice layer to ice cover throughout pancake ice formation and
the mechanics of interaction between the different kinds if ice
\cite{Gosnik83}.

In \cite{Omstedt84} a numerical simulation of ice
formation and transport in turbulent ocean water 
was performed. The upper ocean 
was modelled as an Ekman layer, using a $k-\epsilon$ turbulence model
for the Reynolds-averaged Navier-Stokes (RANS) equations. This model was
improved in \cite{Svensson98} considering the interaction between 
frazil ice particles of
different size. 
In \cite{Jenkins95} a model of frazil ice
transport under an ice shelf was proposed and a
Boussinesq-like approximation was adopted in the limiting case of very small
ice concentration. In this model the vertical structure 
was simplified by assuming all
properties to be constant over the thickness of the plume and 
discontinuous at its boundaries; 
the problem was reduced to one dimension. 
In \cite{Smedsrud04} the model was extended to
the case of ice particles of different size. More recently 
\cite{Skyllingstad01}, a Large Eddy Simulation (LES) 
with a Smagorinsky turbulence model was carried out in
order to study frazil ice dynamics and the ice effects on turbulent 
structures under polar ice covers and in leads. This work took into 
account the effects of salinity.
Finally \cite{Holland05} presented results of LES simulations
for a turbulent flow with frazil ice crystals of different size.

In our work we propose a parametric study of the problem
in the framework of Boussinesq approximation.
In the case of large ice concentrations, the particles
interaction at the ice crystal microscale should be considered. In
particular, it was observed that this phenomenon becomes relevant
for ice volume concentrations above $.003$ \cite{Omstedt84}.
Moreover, the presence of frazil ice crystals produces an increment
in the ice-water mixture effective viscosity. In
\cite{DeCarolis05} it was shown that, for small ice concentrations,
this parameter differs from that of pure fluid by a value of the
same order of the concentration itself.

In this paper we have used a convection model for frazil
ice transport in turbulent water by considering the mixture of ice
crystals and water as a two-phase medium. The balance equations
for mass, momentum and energy of the ice-water mixture are
supplemented by an equation for the temporal evolution of the ice mass
concentration. This equation is strongly coupled to the others. In
particular, the ice particles melting/freezing process and the
temperature fluctuations modify the density of the mixture and
thus the buoyancy force, which, in turn, affects the
hydrodynamical behavior.

The simulation of these equations is subject to
numerical instability due to large round-off errors. In order to
avoid this behavior, related to the large magnitude of the
coefficient of the buoyancy term, we have applied a
Boussinesq-like approximation to the equations of motion.

A complete model of frazil ice dynamics should be able
to represent the complex processes involved in the ice formation,
like growth of ice crystals, flocculation (the clustering of two,
or more, frazil particles), break-up (the cracking of an ice
particle) and secondary nucleation (the production of small ice
crystals by the ice fragments scavenging from crystal surface). As
a consequence, a large range of particles size distribution is
expected with the crystal diameters varying from micrometer to
millimeter. In this paper, however, we have considered only one
particle size. This is not a crude simplification, because the
small ice crystals rapidly evolve into larger ones and finally
only ice particles within a narrow size range actively affect the
system dynamics \cite{Jenkins95}.

For the sake of simplicity we limit our analysis to ice formation
in fresh water and we postpone studying the effect of salinity
to a future work.

We have 
integrated numerically the evolution equations in the Boussinesq limit 
using a 
second order finite difference method.
The smallest scales have been modelled within the 
Large Eddy Simulation approach:
the equations for the resolved scales are obtained 
introducing a spatial filter that removes the
unresolved small scales of turbulence;
the filtered nonlinear terms are modelled using 
the approach of \cite{wong92}.
Numerical simulations have been carried on  to determine the 
sensitivity 
of ice growth rate and saturation process 
to different sets of non-dimensional parameters.

\section[Model equations]{Model equations for frazil
ice transport in turbulent water}\label{Model}

We model the convective motion in the upper ocean mixed-layer as
 convection between two parallel and infinitely long boundaries
(albeit periodic), with the upper boundary corresponding to the
air-ocean interface and the lower one corresponding to the mixed
layer bottom. The distance $L$ between the boundaries
 and the temperature difference $\Delta T$ maintained between them are
set equal, respectively, to a typical depth of the mixed-layer
($30 \div 100$ m.) and to a typical temperature gap between the
air-ocean interface and the mixed-layer bottom ($0.1 \div  0.2
\hspace{0.5mm}^{\circ}$C). The temperature at the
air-ocean interface is set to a value slightly less than the salt water freezing
point $-2.0 \hspace{0.5mm}^{\circ}$C corresponding to a typical
salinity value $S_0 = 34.5$ psu. Salinity variations play only a
small role in the thermodynamics of the system and we shall hold
it constant \cite{Holland05}. Ice concentration is assumed to be
large enough to be represented as a continuum but small enough to
neglect the particles interactions. As a consequence, the system of
water and frazil ice particles can be modelled as a two-component
mixture, which is treated as a locally homogeneous fluid with
averaged properties. In the dilute limit that we are considering,
the effective viscosity of the ice-water mixture differs from that
of pure fluid by a value of the same order of the concentration
itself \cite{DeCarolis05} and this viscosity increment can be
neglected.

We consider a frame of reference such that the $z$
axis is vertically upward.
The ice-water mixture density $\rho$ is given by:

\begin{equation}\label{IceWaterRHO}
\rho = \rho_i \phi_i + \rho_w \phi_w,
\end{equation}

\noindent where $\rho_i$ and $\phi_i$ are, respectively, the ice
density and volume fraction; similarly
 $\rho_w$ and $\phi_w = 1 - \phi_i$ are, respectively, the water density and volume fraction;
 $\rho_w$ is related to the temperature by the following
 equation of state:

\begin{equation}\label{IceWaterRhoW}
\rho_w = \rho_{w0} [1 - \alpha (T - T_0)],
\end{equation}

\noindent where $\rho_{w0}$ is the density corresponding to the
lower boundary temperature $T_0$ and $\alpha$ is the thermal
expansion coefficient.

The total density $\rho$ can be written as:

\begin{equation}\label{IceWaterRhoCi}
\rho = \Big(\frac{C_i}{\rho_i} + \frac{1 - C_i}{\rho_w} \Big)^{-1}
\end{equation}

\noindent in terms of $C_i$, that is the ice mass concentration,
obeying $C_i\rho=\rho_i\phi_i$ (similarly
$C_w = 1 - C_i$ will denote the water mass concentration).

The equation of continuity for the total density $\rho$
is given by:

\begin{equation}\label{IceWaterMass}
 \frac{\partial \rho}{\partial t} + \nabla \cdot (\rho \mathbf{u}) = 0,
\end{equation}

\noindent while the equation for the mixture momentum balance 
is formulated as follows:

\begin{equation}\label{IceWaterMomentum}
\rho \frac{\partial \mathbf{u}}{\partial t}  + \rho (\mathbf{u}
\cdot \nabla) \mathbf{u} = -  \nabla p +  \rho g
 \vec{k} + \nu \rho \nabla^2 \mathbf{u},
\end{equation}

\noindent where $\mathbf{u}$ and $p$ are respectively 
the velocity and the pressure fields, $g$ is the gravity 
acceleration and $\nu$ is the kinematic viscosity.

The equation for the temperature $T$ can be written as follows:

\begin{equation}\label{IceWaterEnergy}
\frac{\partial T}{\partial t} +  \mathbf{u} \cdot \nabla T = k_T
\nabla^2 T + G_T,
\end{equation}

\noindent where $k_T$ is the water thermal diffusivity and $G_T$
is the source term due to melting and freezing processes,
considering the ice particle as a sphere:

\begin{equation}\label{EqDeltaT}
G_T = \frac{\Delta T}{\Delta t} =   \lambda _T  \rho C_i  (T_i -
T),
\end{equation}

\noindent where \cite{holman}:

$$
\lambda_T  =  \frac{3  \overline{h}}{R_S c_p  \rho_w \rho_i}
$$

\noindent and $\overline{h} = \frac{Nu k_w}{2R_S}$ is the heat
transfer
    coefficient ($Nu$ is the Nusselt number and $k_w$ is the sea water
    thermal conductivity, $R_S$ is the ice particle radius), 
    $T_i$ is the ice particle temperature, taken equal to
    the freezing point, and $c_p$ is the water specific heat.

Finally, the equation for the ice component can be written as
(\cite{Hunt69},\cite{Landau59}, \cite{Omstedt84}):

\begin{equation}\label{IceWaterCWRise}
\frac{\partial (\rho C_i)}{\partial t}  +   \nabla \cdot
(\mathbf{u} \rho C_i) = w_r \frac{\partial (\rho C_i)}{\partial
z} + k_i \nabla \cdot (\rho \nabla C_i) + G_i,
\end{equation}

\noindent where $k_i$
is the ice molecular diffusivity, $\mathbf{w_r} = (0,0,w_r)$ is
the rise velocity that derive from the balance between the upward
net buoyancy force and the drag force. It is still an
open question how to model this velocity; we have left it as a
free parameter for sensitivity analysis
(\cite{Adams81},\cite{Hunt69},\cite{Omstedt84},\cite{Svensson98}).
Here $G_i$ is the source term due to melting and freezing
processes and it is given by:

\begin{equation}\label{EqDeltaC}
G_i = \frac{\Delta (\rho C_i)}{\Delta t} =  \lambda _i  \rho C_i
(T_i - T)
\end{equation}

\noindent where \cite{holman}:
$$\lambda _i  =  \frac{3  \overline{h}}{R_S L_i  \rho_i}$$

\noindent and $L_i$ is the ice latent heat of fusion.

It is important to
    remark that we have used spherical particles instead of
    disk-like ones which better represent the frazil ice
    platelets. The choice is motivated by the great simplifications introduced in
    the model. Moreover the ice particle shape affects the source terms only in their magnitude;
    the latter parameters will be used in the foregoing for
    the sensitivity analysis and they will vary in a wide range;
    so we can conclude that the effect of the particle shape on the source terms is not
    significant for the analysis we are going to carry on.

In order to rewrite eqs. (\ref{IceWaterRhoW}-\ref{EqDeltaC})
in non-dimensional form, we consider the depth of the upper ocean
mixed-layer $\bar{L} = L$ as length scale, so the depth of the layer
is 1, the viscous scale
velocity $\bar{U} = \nu / \bar{L}$ as velocity scale, the
corresponding time $\bar{t} = \bar{L} / \bar{U} = \bar{L}^2 / \nu
$ as time scale, the water density $\bar{\rho} = \rho_{w0}$ as
density scale and the temperature gap $\overline{\Delta T}$
between upper and lower boundary as temperature scale. 
We indicate by $z$ the dimensionless vertical 
coordinate with $z_0=-1/2$ the lower boundary.
We obtain:

\begin{equation}\label{IceWaterNonDimRhoW}
\rho_w =  1 - \bar{\alpha} (T - \widehat{T}_0),
\end{equation}

\begin{equation}\label{IceWaterNonDimRho}
\rho = \Big(\frac{C_i}{\rho_i} + \frac{1 - C_i}{\rho_w} \Big)^{-1},
\end{equation}

\begin{equation}\label{IceWaterNonDimMass}
\frac{\partial \rho}{\partial t} + \nabla\cdot (\rho \mathbf{u}) =
0,
\end{equation}

\begin{equation}\label{IceWaterNonDimMomentum}
\rho \frac{\partial \mathbf{u}}{\partial t}  + \rho (\mathbf{u}
\cdot \nabla) \mathbf{u} = -  \nabla p - \frac{Ra}{\bar{\alpha}Pr}
\rho \vec{k} + \rho \nabla^2 \mathbf{u},
\end{equation}

\begin{equation}\label{IceWaterNonDimTemp}
\frac{\partial T}{\partial t} +  \mathbf{u} \cdot \nabla T =
\frac{1}{Pr} \nabla^2 T + \beta \rho C_i (\widehat{T}_i - T),
\end{equation}

\begin{equation}\label{IceWaterNonDimCWRise}
\frac{\partial (\rho C_i)}{\partial t}  +   \nabla \cdot
(\mathbf{u} \rho C_i) = \widehat{w}_r \frac{\partial (\rho
C_i)}{\partial z} +  \frac{1}{Sc} \nabla \cdot (\rho \nabla C_i)
+ \gamma \rho C_i (\widehat{T}_i - T),
\end{equation}

\noindent where $Ra = g \bar{\alpha} \bar{L}^3 / (k \nu)$ is the
Rayleigh number, $Pr = \nu/k$ is the  Prandtl number, $Sc =
\nu/k_i$ is the ice Schmidt number, $\widehat{w}_r = w_r/\bar{U}$,
 $\bar{\alpha} = \alpha \overline{\Delta T}$, $\beta = \frac{\lambda_T \overline{L}^2 \overline{\rho}}{\nu}$,
 $\gamma = \frac{\lambda_i \overline{\Delta T}\hspace{0.5mm}
\overline{L}^2}{\nu}$, $\widehat{T}_i = T_i / \overline{\Delta T}$
and $\widehat{T}_0 = T_0 / \overline{\Delta T}$.

 At the lateral boundaries, we prescribe periodic conditions,
while at the horizontal boundaries we impose a fixed temperature
and free-slip condition for velocity. For frazil ice concentration
we impose Dirichlet conditions $C_i = 0$ 
at the bottom boundary, while the conditions to impose at the
top is still an open question;
we postpone the discussion of such boundary conditions 
at section \ref{deep water} and \ref{free}.

\section[Boussinesq approximation]{The Boussinesq approximation
 for frazil ice transport model}\label{Boussinesq}


 The buoyancy term in the mean momentum balance equation
(\ref{IceWaterNonDimMomentum}) is multiplied by the factor:

$$
\frac{Ra}{\bar{\alpha}Pr}.
$$

In the cases of physical interest, the Rayleigh number is 
in excess of 
$10^{8} \div 10^{9}$ and $\bar{\alpha} \simeq 10^{-6}$; hence this
factor can be estimated to have a magnitude of at least $10^{14} \div
10^{15}$. As a consequence, the numerical evaluation of this term
in a double precision code is subject to significant round-off
errors: this produces severe numerical instability.
In order to
avoid this problem we introduce a Boussinesq-like approximation.
We introduce a hydrostatic equilibrium without ice 
and then we
write down the equations for the small perturbations to this
state. The fluctuations
are written as a power series of the small
parameter $\bar{\alpha}$ and the expansion is truncated to
lowest order. Temperature fluctuations, pressure and velocity field
are of order ${\cal O}(1)$, 
whereas density and ice concentration fluctuations 
are of order ${\cal O}(\bar{\alpha})$.
We thus have:
\[
\mathbf{u} = \mathbf{u'} +{\cal O}(\balpha),\quad
\nabla p = -\frac{Ra}{\balpha Pr}[1+\balpha(z-z_0)]\vec{k}+
\nabla p' + {\cal O}(\balpha),\quad
\]
\[ 
T =\widehat{T}_0 - (z-z_0) + T' +{\cal O}(\balpha),\quad 
C_i=\balpha C'_i+{\cal O}(\balpha^2),\quad
\rho=1+\balpha (z-z_0) +\balpha\rho'+{\cal O}(\balpha^2).
\]
 
\noindent Substituting into eq. (\ref{IceWaterNonDimRho}) 
to leading order in $\balpha$ we obtain:

\begin{equation}\label{IceWaterBousRhoALPHA}
\rho'=-\left(\frac{1-\rho_i}{\rho_i}C'_i+T'\right)
\end{equation}

\noindent and the equations 
(\ref{IceWaterNonDimMass}) and
(\ref{IceWaterNonDimMomentum}) become:
\begin{equation}\label{IceWaterBousMassDisturbEq}
\nabla\cdot \mathbf{u'} = 0,
\end{equation}

\begin{equation}\label{IceWaterBousMomentumDisturbEq}
\frac{\partial \mathbf{u'}}{\partial t}  +  (\mathbf{u'} \cdot
\nabla) \mathbf{u'} = - \nabla p' + \frac{Ra}{Pr}
\left[ \frac{1 - \rho_i}{\rho_i}C_i'  + T' \right] \vec{k} +
\nabla^2 \mathbf{u'}.
\end{equation}

\noindent {{Note that the factor $\frac{Ra}{\bar{\alpha} Pr}$
in the buoyancy term of eq. (\ref{IceWaterNonDimMomentum})
has been transformed to $\frac{Ra}{Pr}$, which is not subject anymore to
numerical instability due to round-off errors.}}

Temperature equation (\ref{IceWaterNonDimTemp}) becomes:
\begin{equation}\label{IceWaterBousTempDisturbEq}
\frac{\partial T'}{\partial t}  + \mathbf{u'} \cdot \nabla T'  =
- w' + \frac{1}{Pr} \nabla^2 T' + \beta \bar{\alpha} C_i'
f (z, T'),
\end{equation}

\noindent where:

$$
f (z, T') = \widehat{T}_i - \widehat{T}_0 - (z-z_0) - T'
$$

\noindent and finally 
the ice transport equation (\ref{IceWaterNonDimCWRise}) reduces
to:

\begin{equation}\label{IceWaterBousCiDisturbEq}
\frac{ \partial C_i'}{\partial t}+ \mathbf{u'} \cdot \nabla C'_i  =
\frac{1}{Sc} \nabla^2 C_i' +  \widehat{w}_r
\frac{\partial C_i'}{\partial z} + \gamma C_i' f(z, T').
\end{equation}

Equations (\ref{IceWaterBousMassDisturbEq}-\ref{IceWaterBousCiDisturbEq}) are 
a complete set of equations for the small
perturbations to the hydrostatic equilibrium.
For sake of simpler notations, we shall drop
from now on the primes from the variable names.

\section{Numerical results}\label{Results}

Equations (\ref{IceWaterBousMassDisturbEq}-\ref{IceWaterBousCiDisturbEq})
were solved numerically 
using a fractional step approach
\cite{le91}
with a third order, three steps, explicit Runge-Kutta scheme.
For the spatial discretization, we have employed
a second order accurate finite difference scheme on a
staggered grid.
For the momentum equation,
each Runge-Kutta time step has been splitted into two steps; the first one
consists in the calculation of a
temporary velocity field without making use of the pressure terms;
the second step is
 a projection of the solution onto a divergence-free space,
which results in
solving a Poisson equation for pressure.
A fast Poisson solver using Fourier and cycling reduction method
was adopted \cite{recipes}.

The effects of the
small unresolved scales on the resolved ones
are modelled by using the Subgrid Scale Model proposed by \cite{wong92}.
We utilize the same
subgrid scale model for the flux of ice concentration 
and for the temperature:
the turbulent Schmidt number 
is computed by using the dynamic procedure
and varies in space and time.

The domain has an extension of $2\pi$ in the horizontal directions
and height equal to one. The resolution varies with Rayleigh number:
at $Ra=10^4, 10^6$ and $10^8$ we used respectively
 $31\times 31\times 30$, $51\times 51\times 36$ and 
$81\times 81\times 60$ collocation points.

\subsection{Ice production in deep water}\label{deep water}

A significant amount of oceanic frazil ice
might be produced in the deep layers of water.
In \cite{Martin81a}, for example, it is stated that 
a large amount of frazil ice
in seawater adjacent to shelves
could be produced
by freezing of parcels of water rising towards
supercooled upper layers.
In this series of simulations, we want to focus on the ice produced
in the regions far from the upper layer. 
To this end we use the boundary conditions $C_i=0$ at $z=\pm 1/2$.
These conditions amount to cancel the convective
flux of ice through the horizontal boundaries. 
They produce a very thin boundary layer immediately beneath
the free surface where a non-physical ice concentration drop is
present. 

In all the subsequent simulations we assume that the freezing temperature
is $\widehat T_i=\widehat T_0+z_0$, 
that is the upper half of the hydrostatic layer is below
freezing point.

\begin{figure}[hbtp]
\begin{center}
\includegraphics[width=10cm,clip=true]{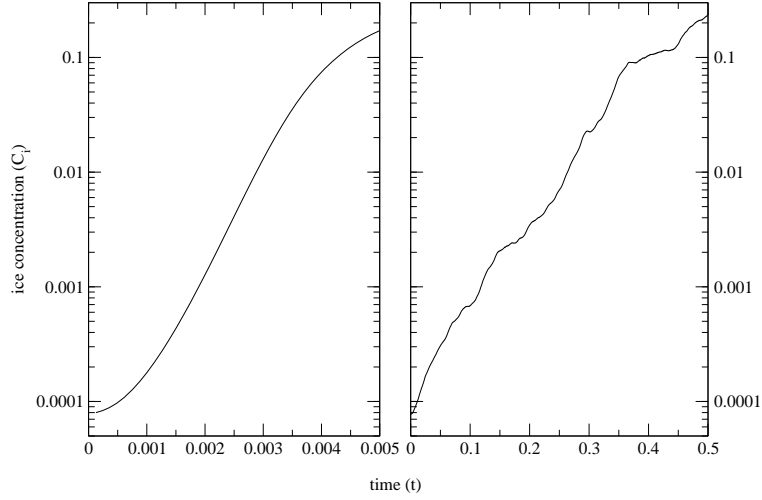}
\caption{Evolution of total ice concentration for 
$Ra=10^6$. Left: $\gamma=10^4$, $w_r=150$. Right:  $\gamma=10^3$, $w_r=10$.
}
\label{fig:C(t)}
\end{center}
\end{figure}

\begin{figure}[hbtp]
\begin{center}
\includegraphics[width=10cm,clip=true]{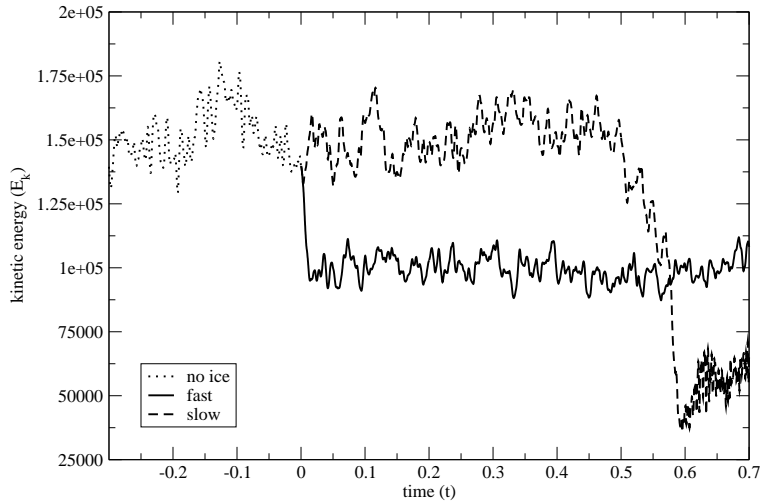}
\caption{Evolution of kinetic energy for the two 
simulations of Fig. \ref{fig:C(t)}. Ice seeds
are introduced at time $t=0$.
Full line: $\gamma=10^4$, $w_r=150$ (fast growth).
Broken line: $\gamma=10^3$, $w_r=10$ (slow growth).
}
\label{fig:ener_fast_slow}
\end{center}
\end{figure}

Typical curves of the total ice concentration versus time 
are shown in Fig. \ref{fig:C(t)}. In the left figure,
we clearly distinguish a short transient followed by an exponential
growth,  ended by the nonlinear saturation.
Evolution of kinetic energy is shown in Fig. 
\ref{fig:ener_fast_slow}. The ice seeds are introduced at time $t=0$.
We observe that the production of ice
is fast compared to the large eddies turnover time.
As a consequence,
during the linear phase, the velocity and temperature fields 
are almost constant and 
we get an almost pure exponential growth of ice concentration.
In the right figure, instead, the exponential growth phase does not
appear very clearly. This is because the time scale for the ice formation
is large compared to the convection turnover time:
the flow 
changes significantly during ice formation. 
The fast and slow regimes appear very clearly in Fig. \ref{fig:ener_fast_slow}.

%
%

\begin{figure}[hbtp]
\begin{center}
\includegraphics[width=10cm,clip=true]{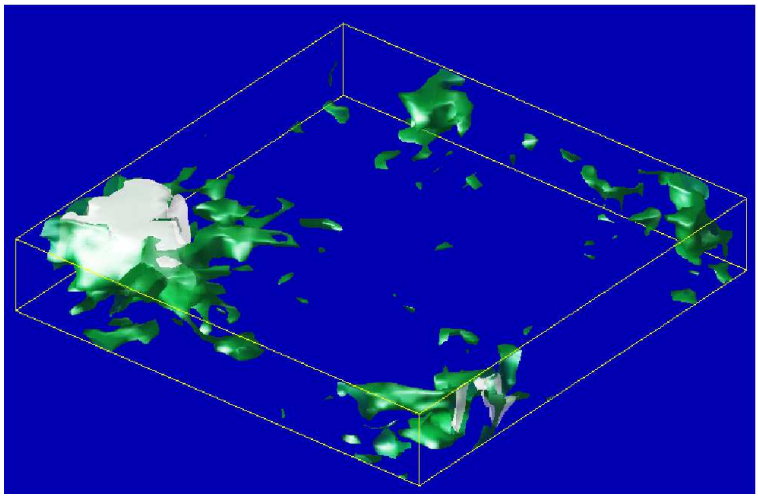}
\caption{Isosurface of negative temperature fluctuations (-0.2, green)
and isosurfaces of large ice concentration ($.1 C_{\max}$, white).
Snapshot taken at a temporal station during the exponential growth regime
for a fast ice formation regime
($Ra=10^6$, $\gamma=10^4$, $w_r=150$).}
\label{fig:iso-fast}
\end{center}
\end{figure}

\begin{figure}[hbtp]
\begin{center}
\includegraphics[width=10cm,clip=true]{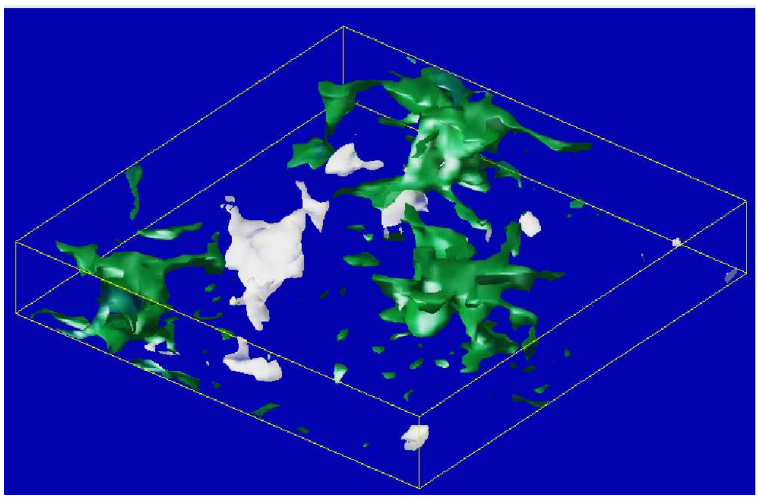}
\caption{Isosurface of negative temperature fluctuations (-0.2, green) 
and isosurfaces of large ice concentration ($.1 C_{\max}$, white)
for a slow ice formation regime
($Ra=10^6$, $\gamma=10^3$, $w_r=10$)}
\label{fig:iso-slow}
\end{center}
\end{figure}

In the fast growth regime, we observe that most of the ice 
is produced inside the strongest downwelling plumes.
A typical situation is shown in Fig. \ref{fig:iso-fast}:
it is a snapshot of the flow taken during the linear phase of ice growth.
The isovalue for temperature fluctuation has been chosen
so that only the most intense downwelling plume appears
(since the flow is periodic in horizontal directions the four 
large green structures are actually contiguous).
We see that ice (white structures) is present mostly at the top of that plume. 
This is natural, since it is also
the coldest place inside the flow:
the largest part of ice is produced there, and,
since the time scale for the linear phase is small, the shape of the plume
almost does not change and the ice remains confined inside it.
By contrast, in the slow regime, the spots of large ice concentration
do not necessarily reside in the large downwelling plumes.
Fig. \ref{fig:iso-slow} shows a typical situation where we
see small spots of ice next to the strongest downwelling plumes,
and a large spot in a relatively quiescent region in the flow.
The ice still forms mostly inside the 
coldest downward plumes, but now the process of ice formation is
slow in comparison to the convection time scales and the ice forming
has time to migrate towards quiescent regions. 
Looking again at Fig. \ref{fig:ener_fast_slow},
we remark that the total
kinetic energy decreases as soon as a significant amount of ice is produced;
this happens because most of the kinetic energy is concentrated
in the most intense plumes: the ice that is produced in
the strong downwelling plumes modifies the buoyancy force and 
tends to slow down the downward motion; as a consequence,
the kinetic energy of those plumes is reduced.

In Figs. \ref{fig:omega(wr).ra1e6} and \ref{fig:omega(gamma).ra1e6}
we plot the growth rate for the linear phase as a function
respectively of the rise velocity $w_r$ and of the 
thermal exchange coefficient $\gamma$.
Each point in the figures is obtained by carrying on a full simulation 
until nonlinear saturation of ice concentration occurs. 
In the fast regime, the flow is almost frozen during the 
ice production and the growth is almost exponential; 
the growth rate is computed as the slope of ice concentration
versus time in log scale (as in  Fig. \ref{fig:C(t)} (left)),
without considering the initial transient and the final nonlinear regime.
In the slow regime the value of the growth rate is determined as the 
slope of the line 
which fits at best the curve in the linear phase, omitting the 
initial transient.

We notice in Fig. \ref{fig:omega(wr).ra1e6} 
that the growth rate has a maximum at a particular value of the 
rise velocity which we denote $\wmax$. For $Ra=10^6$ 
it is about $\wmax \simeq 75$.

\begin{figure}[hbtp]
\begin{center}
\includegraphics[width=10cm,clip=true]{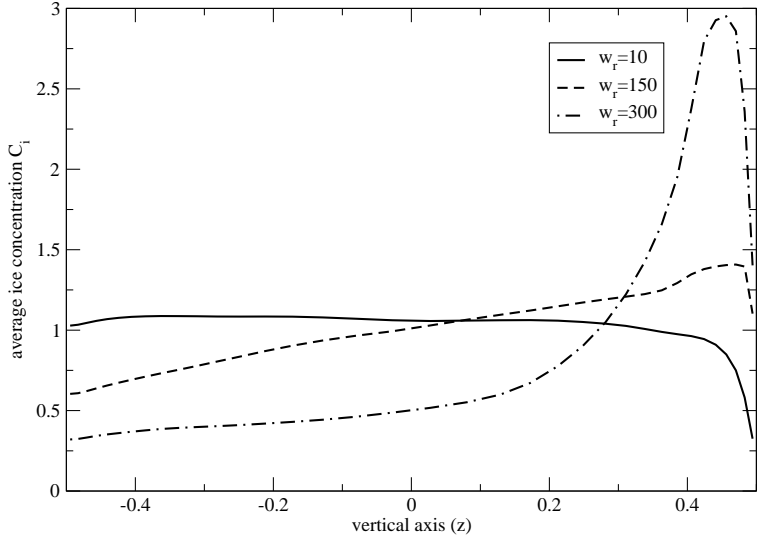}
\caption{Average ice concentration for different values of $w_r$
($Ra=10^6,\gamma=10^3$)}
\label{fig:specmed_wr}
\end{center}
\end{figure}

Looking at the concentration profiles averaged in 
the horizontal directions (Fig. \ref{fig:specmed_wr}) 
we can distinguish three situations according to the value of $w_r$.

In the large rise velocity regime ($w_r>\wmax $),
the ice crystals are rapidly pushed
upward and they cannot act as nuclei for ice formation;
therefore the ice rate of growth decreases when
rise velocity increases and most of the ice
concentration is found near the top.
If $w_r\sim \wmax $, the rise velocity is balanced by the 
downwelling velocity of the plumes in the upper layer of the domain, where
the temperature is the lowest.
Crystals remain confined in this region and their concentration grows rapidly.
Finally if $w_r<\wmax $ the ice 
is transported down and exits the plumes where it was created. 
The ice spreads throughout the domain and the average concentration profile
is almost flat (Fig. \ref{fig:specmed_wr}).

Note that the profiles 
observed by (\cite{Pegau96})
beneath lead in the Beaufort Sea (north of Alaska) and those
obtained by (\cite{Skyllingstad01}) from the numerical
simulations of lead dynamics show a 
large presence of ice near the top of the layer; this is 
in agreement with the situation where $w_r\gtrsim \wmax $.

\begin{figure}[hbtp]
\begin{minipage}[t]{0.45\linewidth}
\centerline{\includegraphics[width=\linewidth,clip=true]{ra1e6_omegawr.eps}}
\caption{Growth rate of ice concentration versus rise velocity for
         $Ra=10^6$ and $\gamma=10^4$.
}
\label{fig:omega(wr).ra1e6}
\end{minipage}
\hfill
\begin{minipage}[t]{0.45\linewidth}
\centerline{\includegraphics[width=\linewidth,clip=true]{ra1e6_omegagamma.eps}}
\caption{Growth rate of ice concentration versus $\gamma$ for
         $Ra=10^6$ and $w_r=150$.
}
\label{fig:omega(gamma).ra1e6}
\end{minipage}
\end{figure}

\subsection{A 1D model for fast ice formation}\label{subs:1D model}

We propose in this subsection a simple model 
that explains the behavior of the
curves shown in Figs. \ref{fig:omega(wr).ra1e6}
 and \ref{fig:omega(gamma).ra1e6}
for the fast regime.
We start from (\ref{IceWaterBousCiDisturbEq}) and 
assume that inside the intense vertical plumes we can neglect
horizontal velocities and horizontal derivatives.
Therefore $C_i$ is a function of $z$ and $t$ and we get:

\begin{equation} \label{eq:ice 1D}
\frac{\partial C_i}{\partial t}= -W\frac{\partial C_i}{\partial z}
+\frac{1}{S}\frac{\partial^2C_i}{\partial z^2}+\Gamma C_i,
\end{equation}
where:
\beq
\frac{1}{S}=\frac{1}{Sc}+\frac{1}{Sc_{\rm turb}},\quad
W=w+w_r,\quad
\Gamma=\gamma(T_i-T).
\label{eq:SWG}
\eeq
We perform an analysis into normal modes:
\[
C_i(t,z)=\widehat C (z)e^{\omega t}
\]
where \noindent $\omega \in \mathbb{C}$ is the eigenvalue and $\widehat C (z)$ 
the eigenvector:
\beq
\widehat C''(z)-WS\widehat C'(z)+
(\Gamma-\omega)S\widehat C (z)=0
\label{eq:ice 1D eig}
\eeq
($\widehat C'$ and $\widehat C''$ denote 
derivatives with respect to $z$).
Let us further assume that $W$, $\Gamma$ and $S$ are constant.
We look for solutions with exponential dependence on $z$.
Imposing the boundary conditions:
\[
\widehat C(-1/2) = \widehat C(1/2)=0,
\]
we get:
\[
\omega=\Gamma-\frac{W^2S}{4}- \frac{n^2\pi^2}{S}\, \qquad |n|\neq 0
\]

$$C_i(z,t)=ae^{\omega t}e^{\frac{WSz}{2}}
\left([1-(-1)^n]\cos{n\pi z}+i[1+(-1)^n]\sin{n\pi z}\right),
$$
where $n$ is a non vanishing integer 
and $a$ is an arbitrary constant.
The case $n=0$ is degenerate;
the growth rate is:
\beq
\omega=\Gamma-\frac{W^2S}{4}.
\label{eq:omega}
\eeq
We can now impose only one boundary condition.
If $W>0$ the eigenvector is:
\[
\widehat C(z)=ae^{\frac{WSz}{2}}(1-2z)
\]
and the boundary condition is exact in $z=1/2$ and
only approximately zero in $z=-1/2$.
If $W<0$:
\[
\widehat C(z)=ae^{\frac{WSz}{2}}(1+2z)
\]
and the boundary condition is exact in $z=-1/2$ and
only approximately zero in $z=1/2$.

\begin{figure}[hbtp]
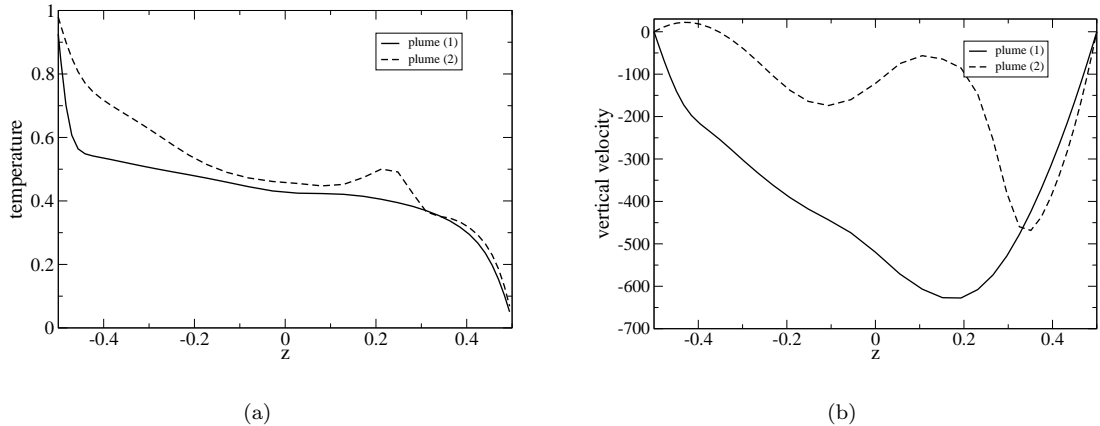

   \subfigure[]{
\label{fig:plume_T}
\begin{minipage}[b]{0.5\linewidth}
\centering \includegraphics[width=.9\linewidth,clip=true]{plume_T.eps}
\end{minipage}
}
\hfill
   \subfigure[]{
\label{fig:plume_w}
\begin{minipage}[b]{0.5\linewidth}
\centering \includegraphics[width=.9\linewidth,clip=true]{plume_w.eps}
\end{minipage}
}
\caption{
Vertical profile of temperature fluctuations (a) and vertical velocity
(b) of the two most most intense
downwelling plume. Data are averaged over the horizontal 
dimension of the plume. $Ra=10^6$, $\gamma=10^4$, $w_r=150$.
}
\label{fig:profili}
\end{figure}

\begin{figure}[hbtp]
\begin{center}
\includegraphics[width=10cm,clip=true]{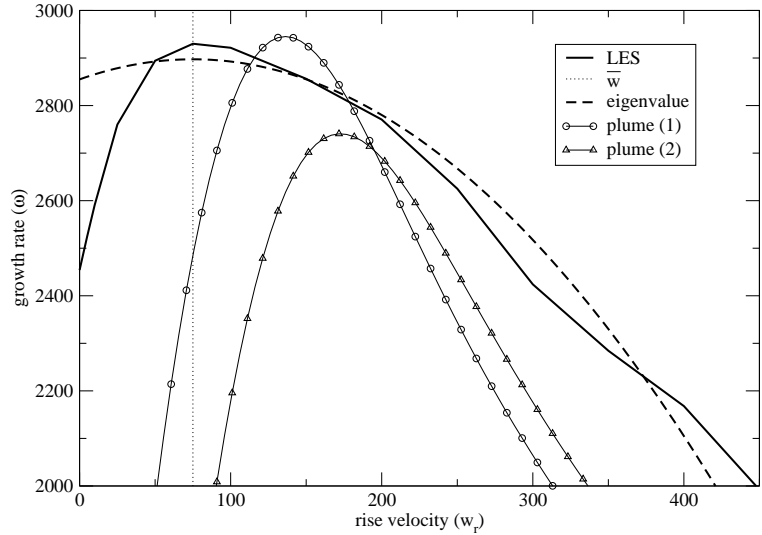}
\caption[]{Growth rate of ice concentration versus rise velocity for 
$\gamma=10^4$ and $Ra=10^6$. \\
Full line: Growth rate from LES simulations.\\
Broken line: eigenvalue given by (\ref{eq:omega}). 
\\
Line with circles: eigenvalue of problem (\ref{eq:eigenproblem})
for the most intense downwelling plume.\\
Line with triangles: eigenvalue of problem (\ref{eq:eigenproblem})
for the second most intense downwelling plume
}
\label{fig:wrise_ra6}
\end{center}
\end{figure}

\begin{figure}[hbtp]
\begin{center}
\includegraphics[width=10cm,clip=true]{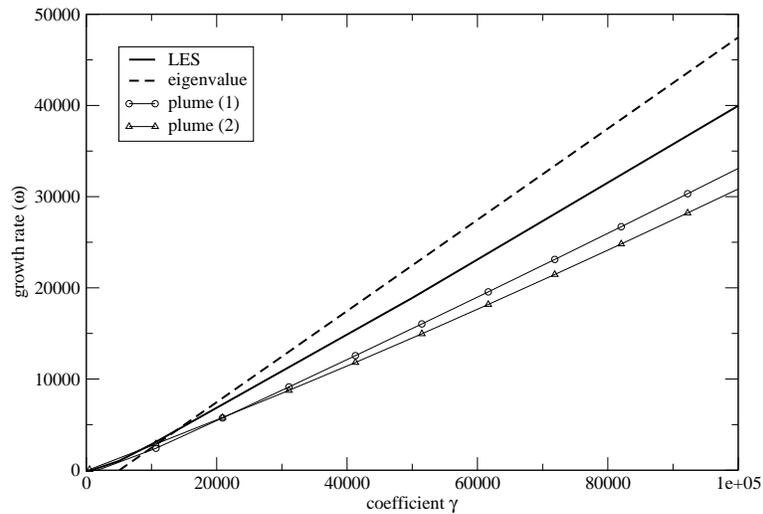}
\caption{Growth rate of ice concentration versus $\gamma$ for 
$w_r=150$ and $Ra=10^6$. For the legend see 
Fig \ref{fig:wrise_ra6}.
}
\label{fig:gamma_ra6}
\end{center}
\end{figure}

The degenerate case is the one with the largest growth rate 
and it is therefore the most interesting. 
In Figs. \ref{fig:wrise_ra6} and \ref{fig:gamma_ra6}
we plot the growth rate $\omega$ versus respectively $w_r$ and $\Gamma$
 together with the actual values obtained by the LES
simulations. The values of $W$, $\Gamma$ and $S$ are expressed by 
(\ref{eq:SWG}) and depend on typical values of vertical velocity $w$,
temperature $T$ and turbulent Schmidt number $Sc_{\rm turb}$.
Those values could be predicted by turbulent convection theories but we
decide here to choose the values which fit at best the LES curve. As we
see in Fig. \ref{fig:wrise_ra6}, the fit is very good for rise velocities
greater or equal to $\wmax$. The bad behaviour at smaller values
is due to the fact that thei ice does not stay confined inside downwelling plumes
and therefore the hypotheis of our model is not verified.
In order to get a more quantitative agreement, we refine
the model by dropping the assumption that 
$W$ and $\Gamma$ are constant. 
We compute the actual vertical profile of these functions 
by averaging the values obtained by the LES
simulations inside the most intense downwelling plume 
(see Figs. \ref{fig:profili}).

The differential eigenvalue problem becomes:
\begin{equation}
\left\{ \begin{array}{l}  \frac{\sd 1}{\sd S} \widehat C''-W(z) \widehat C'
+\Gamma(z) \widehat C=\omega \widehat C,\\ \\
\widehat C(-1/2)=\widehat C(1/2)=0,
\end{array} \right.
\label{eq:eigenproblem}
\end{equation}
for prescribed functions $W(z)$ and $\Gamma(z)$.
The solution is no more analytical and we compute it numerically.
We adopt a spectral technique: we expand the solution in series
of truncated Chebyshev polynomials and  
we discretize the problem over the Gauss-Lobatto grid points:
\[
z_j=\frac{1}{2}\cos{\frac{\pi j}{N}}\qquad j=0,\ldots,N.
\]
We obtain a linear algebraic generalized eigenvalue problem:
$$A\widehat C=\omega B\widehat C,$$
where $A$ and $B$ are $(N\!+\!1)\times(N\!+\!1)$ matrices.
A method to solve such problem is to use the QZ algorithm.
Such algorithm gives the full spectrum
of eigenvalues/eigenvectors.
Another method is to 
use iterative techniques
(Krylov based methods), such as the Arnoldi method or the unsymmetric
Lanczos method \cite{CHA.93}. This method gives few
eigenvalues corresponding to the least stable part of the spectrum.
For our problem we have implemented both the QZ and the Arnoldi method and
we have checked that the results produced by both techniques
are always similar.

The growth rates obtained by solving
(\ref{eq:eigenproblem}) with the vertical profiles of 
the two most intense downwelling plumes and 
those given by the constant 
coefficient model (\ref{eq:omega})
are shown in Figs. \ref{fig:wrise_ra6} and \ref{fig:gamma_ra6}.
From Fig. \ref{fig:wrise_ra6}
we observe that when the rise velocity is far from the 
peak value $\wmax$,
the growth rate predicted by our model underestimates the actual value
(this occurs for about $w_r > 300 $ and $w_r<75$).
The reason is that when the  growth rate is small
the flow cannot be considered steady during the 
exponential ice growth, but rather it fluctuates. 

\begin{figure}[hbtp]
\begin{center}
\includegraphics[width=10cm,clip=true]{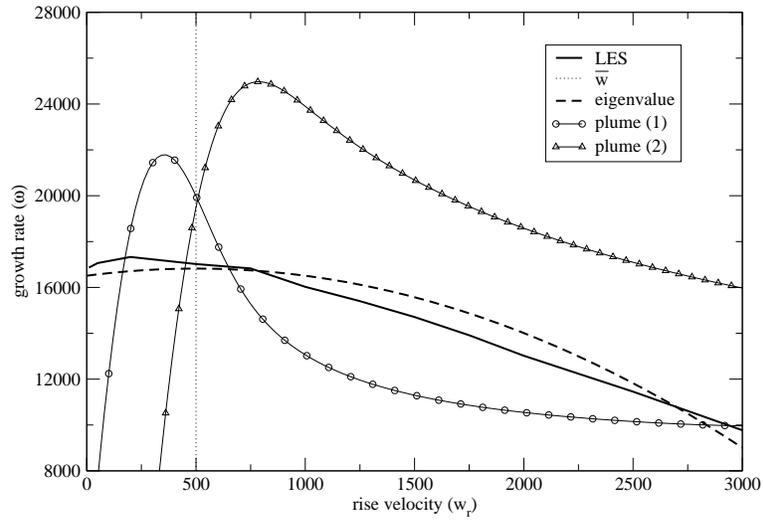}
\caption{Growth rate of ice concentration versus rise velocity for 
        $\gamma=10^5$ and $Ra=10^8$}
\label{fig:wrise_ra8}
\end{center}
\end{figure}

\begin{figure}[hbtp]
\begin{center}
\includegraphics[width=10cm,clip=true]{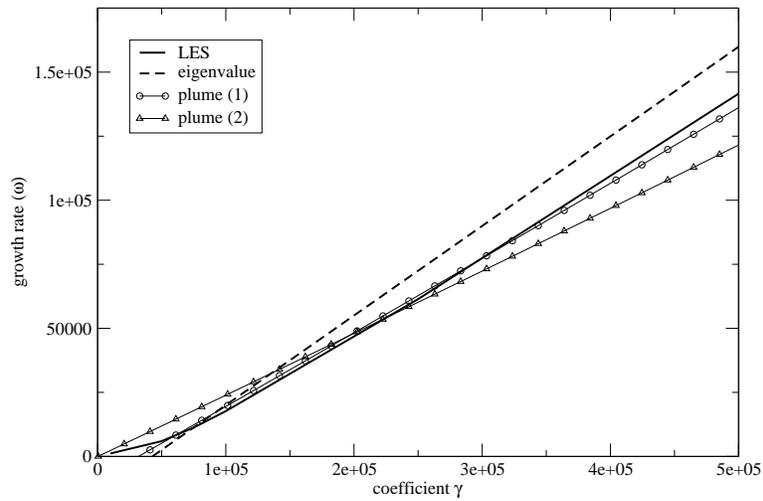}
\caption{Growth rate of ice concentration versus $\gamma$ for 
        $w_r=1000$ and $Ra=10^8$}
\label{fig:gamma_ra8}
\end{center}
\end{figure}

Figures  \ref{fig:wrise_ra8} and \ref{fig:gamma_ra8}
are similar to Figs.  \ref{fig:wrise_ra6} and \ref{fig:gamma_ra6},
except for the Rayleigh number that is $10^8$ instead of $10^6$.
We see that the results obtained at $Ra=10^6$
are still valid, thus they are robust and we can
hope to extrapolate them to even larger Rayleigh numbers that are
of interest for geophysical applications.


Some analytical progress far from the fast regime
is possible provided $w_r$ is not too small (so to be able to 
balance the downwelling in the plumes). 
In this way, the permanence time $\tau_p$
of an ice crystal in a plume can be considered long and the growth process will be
dominated by the plumes with the largest growth rate:
$ \langle\exp(\Gamma\tau_p)\rangle\sim\exp(\Gamma_{\scriptscriptstyle\rm MAX}\tau_p)$.
Taking at the same time the limit $\Gamma\tau\ll 1$ with $\tau$ the typical
plume circulation time, allows to treat the velocity and temperature
fluctuations as a noise and to calculate averages in closed form.

Let us look at a single Lagrangian parcel,
moving with velocity $w(z(t),t)+w_r$ (plus the
effect of molecular diffusion) and
endowed with an ice concentration $c(t)$ obeying:
$$
 \frac{\d c(t)}{\d t}=\gamma [T_i-T(z(t),t)]c(t).
$$
Identify with tilde and overbar, the fluctuating and mean field component
of $w$ and $T$ at the parcel position $z$:
$w(z,t)=\bar w(z,t)+\tilde w(t)$;
$T(z,t)=\bar T(z,t)+\tilde T(t)$
In the limit $\tau\to 0$, the equations for
$z(t)$ and $c(t)$ can be written in the form:
\begin{displaymath}
\left\{
\begin{array}{l}
\d z=[w_r+\bar w(z,t)]\d t+\sigma_{\tilde w}\tau^{1/2}\d B_1+(2/Sc)^{1/2}\d B_2,\\
\d c=\gamma[T_i-\bar T(z,t)]c\,\d t+\gamma\sigma_{\tilde T}\tau^{1/2}c\,\d^\smalS B_3,
\end{array} \right.
\label{pier2}
\end{displaymath}
where the $\d B_i$ are Brownian increments, that, for simplicity, we take as independent:
$\langle\d B_i\rangle=0$, $\langle\d B_i\d B_j\rangle=\delta_{ij}\d t$.
The superscript in  $\d^\smalS B_3$ indicates Stratonovich prescription
\cite{gardiner}, and arises from the short correlation time limit
$\int_t^{t+\tau} c(t')\tilde T(t')\d t'
\simeq\frac{1}{2}\sigma_{\tilde T}\tau^{1/2}[c(t)+c(t+\tau)][B_3(t)-B_3(t+\tau)]$.
This leads to a correction to the mean growth rate, and the second of eq. (\ref{pier2})
will take the form, back to the standard It\^o prescription:
$$
\d c=\gamma[T_i-\bar T(z,t)+\frac{1}{2}\gamma\sigma_{\tilde T}^2\tau]c\,\d t
+\gamma\sigma_{\tilde T}\tau^{1/2}c\,\d B_3.
$$
The PDF $\rho(z,c;t)$ will then obey the Fokker-Planck equation:
\beq
\frac{\partial\rho}{\partial t}+\frac{\partial}{\partial z}[(w_r+\bar w)\rho]+
\gamma\frac{\partial}{\partial c}[(T_i-\bar T+\frac{1}{2}\gamma\sigma_{\tilde T}^2\tau)c\rho]
=
\frac{\partial^2}{\partial z}(\rho/\bar S)
+\frac{1}{2}\frac{\partial^2}{\partial c}(\gamma\sigma_{\tilde T}^2\tau c^2\rho),
\label{pier4}
\eeq
where $\bar S^{-1}=Sc_0^{-1}+\sigma_{\tilde w}^2\tau/2$ may include non-subgrid
contributions to the fluctuations. The concentration field resulting from
this averaging procedure will be: $\bar C_i(z,t)=
\int\rho(c,z;t)\, c\d c$, and, substituting into
eq. (\ref{pier4}), we obtain:
\beq
\frac{\partial\bar C_i}{\partial t}+\frac{\partial}{\partial z}
[(w_r+\bar w)\bar C_i]=\frac{\partial^2}{\partial z}\bar C_i/\bar S+
\gamma(T_i-\bar T+\frac{1}{2}\gamma\sigma_{\tilde T}^2\tau)\bar C_i.
\label{pier5}
\eeq
Thus, the temperature fluctuations lead to a net increase in the
ice concentration growth rate:
\beq
\Gamma\to
\Gamma+\frac{1}{2}\gamma\sigma_{\tilde T}^2\tau.
\label{pier6}
\eeq
Neglecting this effect, as shown in Fig. \ref{fig:wrise_ra6},
leads to underestimating the actual growth rate.
Similarly, in Fig. \ref{fig:gamma_ra6}
the slope of theoretical curves should be increased, which
is precisely what eq. (\ref{pier6}) does.

Notice that Eq. (\ref{eq:ice 1D}) would be recovered from (\ref{pier5}) (as it should)
if no fluctuating component were singled out of the $w$ and $T$ fields.

\subsection{Non linear phase}
The nonlinear saturation occurs when the ice concentration 
(rescaled by the factor $\alpha$, see (\ref{IceWaterBousCiDisturbEq}))
reaches values of the order of unity: the density of the mixture 
varies by amounts of order $\alpha$ and the buoyancy force
in (\ref{IceWaterBousMomentumDisturbEq}) is significantly affected.
One must bear in mind that the Boussinesq approximation breaks
down when ice concentration exceeds values of order unity; therefore
results of the nonlinear stage must be taken with caution:
they give qualitative indications but they cannot be 
used to assert quantitative conclusions.

\begin{figure}[hbtp]
\begin{center}
\includegraphics[width=10cm,clip=true]{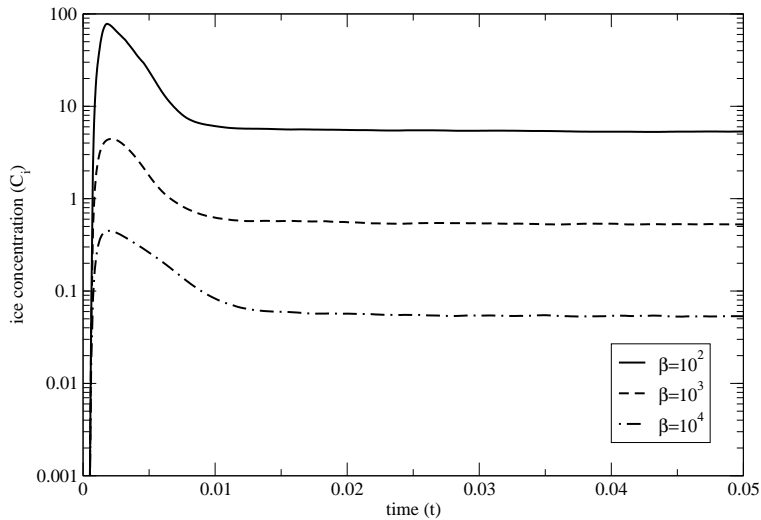}
\caption{Evolution of ice concentration for different values of 
         $\beta$ for $Ra=10^6$, $w_r=100$ and $\gamma=10^5$}
\label{fig:nonlinear_ice}
\end{center}
\end{figure}

In Fig. \ref{fig:nonlinear_ice} we plot the temporal 
evolution of total ice
concentration for three simulations with different values of $\beta$.
The linear stage is insensitive to the value of $\beta$ 
because the ice concentration is too low to produce 
any change in temperature fluctuations.
However we see that the nonlinear saturation is affected
by the value of $\beta$ and the general trend is that 
for larger values of $\beta$ 
the ice saturation levels tend to decrease.

\begin{figure}[hbtp]
\begin{center}
\includegraphics[width=10cm,clip=true]{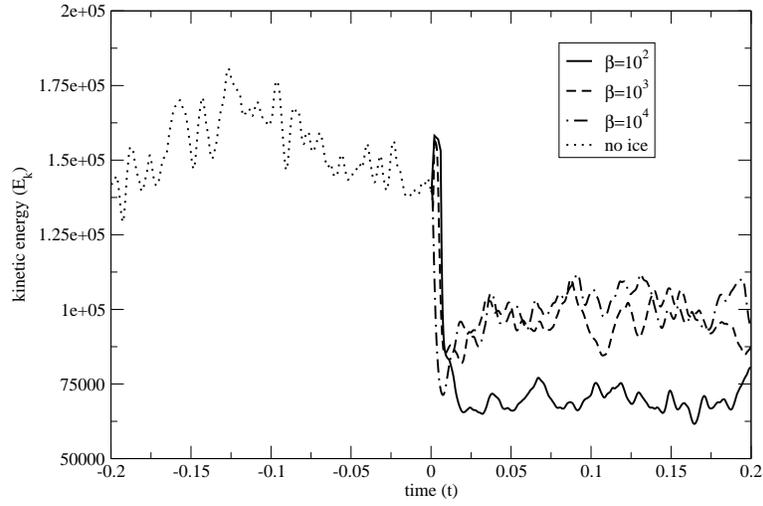}
\caption{Evolution of kinetic energy for different values of 
         $\beta$ for $Ra=10^6$, $w_r=100$ and $\gamma=10^5$}
\label{fig:nonlinear_ener}
\end{center}
\end{figure}

In Fig. \ref{fig:nonlinear_ener}
we see that the presence of ice inhibits convection.

\begin{figure}[hbtp]
\begin{center}
\includegraphics[width=10cm,clip=true]{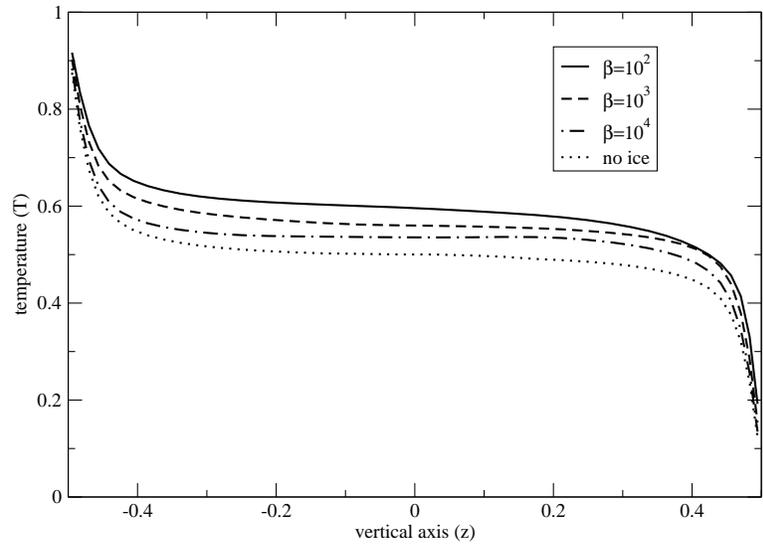}
\caption{Mean vertical temperature for different values of 
         $\beta$ for $Ra=10^6$, $w_r=100$ and $\gamma=10^5$}
\label{fig:nonlinear_temp}
\end{center}
\end{figure}

In Fig. \ref{fig:nonlinear_temp}
we plot the average vertical temperature profiles 
during the nonlinear stage. We see that temperature tends to increase 
when $\beta$ decreases. Lower values of $\beta$ 
correspond to larger values of ice concentration, as we saw
in Fig. \ref{fig:nonlinear_ice}. Therefore 
when total ice concentration increases the temperature of the 
flow tends to increase.
This can be explained by the fact that during the phase transition in 
the freezing process there is an emission of heat. Therefore
more ice production corresponds to more heat in the flow and the 
average temperature tends to increase.

\subsection{Free surface flow}\label{free}

In  subsection \ref{deep water}
we were interested in the ice produced
in the regions far from the upper layer, and therefore we adopted the simple boundary condition
$C_i=0$ for the ice concentration at the surface.

In this subsection we want to study if and how the 
upper boundary conditions affect the ice concentration profile and growth.
There is still no general consensus on 
how to obtain
no upward ice flux at the air-water interface. Some authors
(\cite{Omstedt84}, \cite{Omstedt85} and \cite{Svensson98})
 impose Neumann condition $\partial C_i /
\partial z = h_c$, with $h_c \neq 0$, in order to represent the seeding effect due to
the mass exchange with the atmosphere. It
is still not clear, however, how to model the flux $h_c$.  Other
authors \cite{Skyllingstad01} propose to account for the ice
crystals escaping across the air-water interface by accumulating
them in a floating ice slab. 
Here we consider the situation where 
no ice sheet has formed at the water surface, therefore the 
water surface is a free surface, and we neglect the seeding effect. 
The suitable boundary conditions for this situation are 
free slip conditions for velocity and total vanishing flux for 
ice concentration: 
\[
w=0,\quad
\frac{\partial u}{\partial z}=
\frac{\partial v}{\partial z}= 0,\quad
\frac{\partial C_i}{\partial z}-w_r S C_i= 0
\]
where $S$ is the effective Schmidt number:
$\frac{1}{S}=\frac{1}{Sc}+\frac{1}{Sc_{\rm turb}}$.
The only difference with subsection \ref{deep water} is that 
we impose vanishing total ice concentration 
flux instead of vanishing convective flux.

\begin{figure}[hbtp]
\begin{center}
\includegraphics[width=10cm,clip=true]{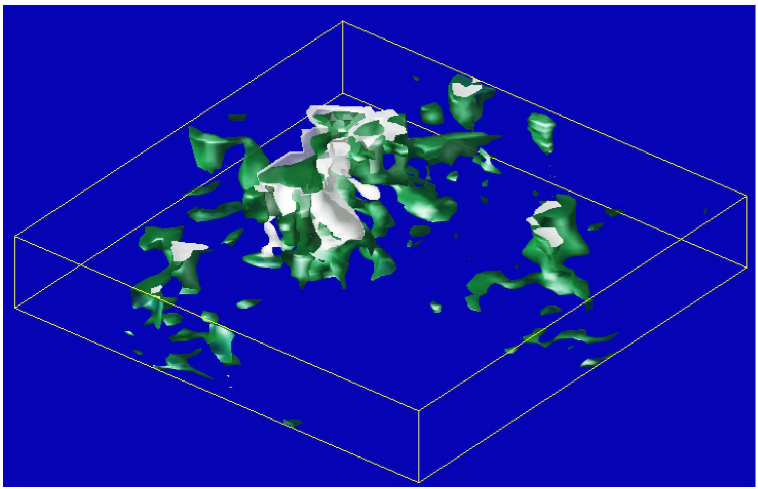}
\caption{Isosurface of negative temperature fluctuations (-0.2, green)
and isosurfaces of large ice concentration ($.1 C_{\max}$, white).
Snapshot taken at a temporal station during the exponential growth regime
for a fast ice formation regime
($Ra=10^6$, $\gamma=10^4$, $w_r=150$).}
\label{fig:newcc_snap}
\end{center}
\end{figure}

\begin{figure}[hbtp]
\begin{center}
\includegraphics[width=10cm,clip=true]{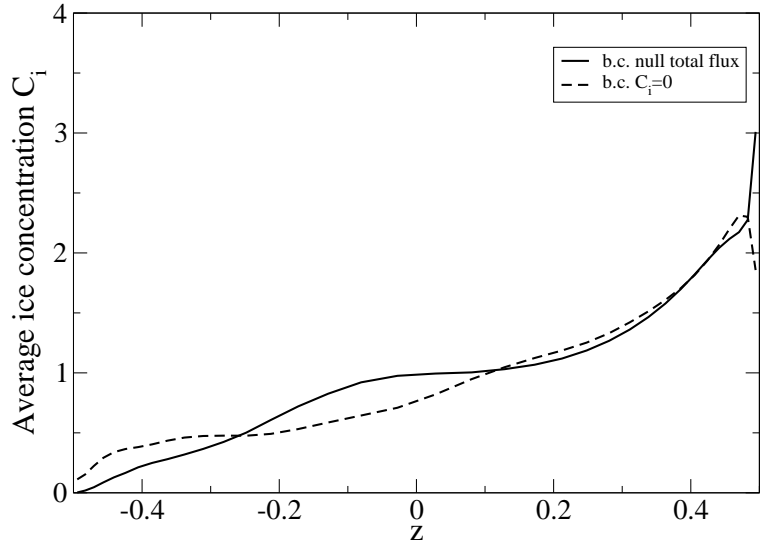}
\caption{Vertical profile of ice concentration averaged over
the horizontal directions at a temporal frame
during the exponential growth regime. The two curves
differ by the boundary conditions adopted at the water-air interface.
The parameters are $Ra=10^6$, $\gamma=10^4$, $w_r=150$.
}
\label{fig:confronti_spec2}
\end{center}
\end{figure}

In Fig. \ref{fig:confronti_spec2} we plot the 
vertical profile of ice concentration averaged over
horizontal directions; we also plot
the one obtained using the boundary
conditions of  subsection \ref{deep water}. 
We see that  
the two curves differ significantly near the surface,
where the new boundary conditions
produce a peak of ice concentration.
Far from the upper boundary however the two profiles look
very similar.
In Fig. \ref{fig:newcc_snap} 
we present 
a snapshot taken during the linear phase 
of the large downwelling plumes and of the ice concentration.
Comparison with Fig. \ref{fig:iso-fast} confirms that the
main difference is the accumulation of ice
near the water surface.

\begin{figure}[hbtp]
\begin{center}
\includegraphics[width=10cm,clip=true]{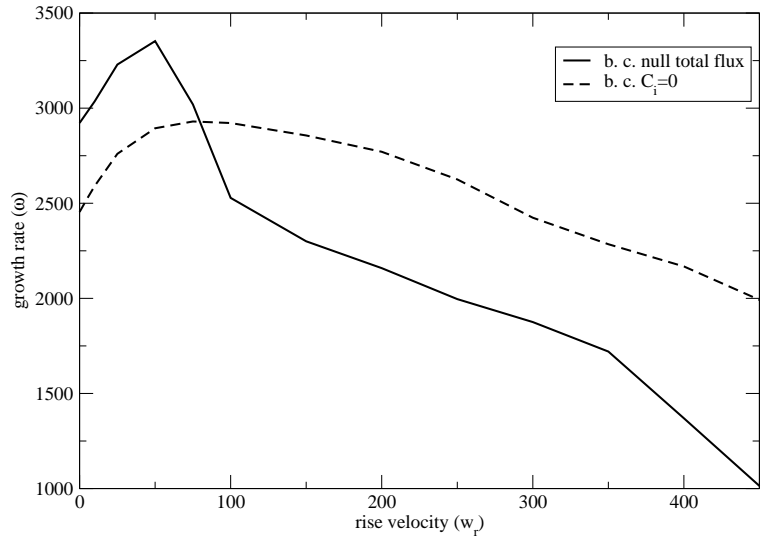}
\caption{Growth rate of ice concentration versus rise velocity for 
different boundary conditions ($Ra=10^6,\gamma=10^4$)
}
\label{fig:newcc}
\end{center}
\end{figure}

We compare next the growth rate of ice concentration for the 
new and old boundary conditions:
in Fig. \ref{fig:newcc} 
we present the growth rate of ice concentration as a function
of the rise velocity $w_r$.            
We observe that the two curves are similar.
When the rise velocity $w_r$ is small, we observe that
the growth rate using the new boundary conditions 
is larger. This can be explained by the fact
that these boundary conditions allow production of ice near
the surface. 

We can thus conclude that the new boundary conditions 
only affect the ice produced 
at the top of the layer, but they do not affect significantly the overall
ice production rate and the ice concentration profile in the deep
layers.

\section{Conclusions}\label{Conclusions}

In this paper we have explored the sensitivity of the
ice formation process to a set of non-dimensional parameters: the
ice melting/freezing source term $\gamma$, the ratio between the
latter and the temperature source term $\beta$, and the ice
particle rise velocity $\widehat{w}_r$. Regarding the Rayleigh
number $Ra$, we have considered moderately turbulent regimes
up to $Ra = 10^8$.

We have modeled the mixture of ice crystals and water
as a two-phase medium composed of water and ice particles of fixed
diameter. The equations describing the system are the classical
balance equations for mass, momentum and energy in the ice-water
mixture, supplemented by an advection-diffusion equation for the
ice mass concentration with an additional source term, which
account for the net ice production during melting or freezing
process. In order to account for the net heat exchange during the
same process, an additional source term has been introduced in the
temperature equation. The simulation of the complete equations,
however, is subject to numerical instability due to large round-off
errors. The origin of such error is the factor $Ra / \bar{\alpha}$
($\bar{\alpha}$ is the coefficient of thermal expansion) in the
buoyancy term, which, for values of $Ra$ of interest ($Ra 
\geq  
10^{8}$), is larger than $10^{14}$. Therefore, we have developed and 
applied
the Boussinesq approximation and we have 
integrated numerically the set of equations 
making use of a numerical code
based on second order finite difference.
A dynamic LES model has been used in order to cope with
the subgrid scales that are not considered in the simulations.

The sensitivity analysis has shown that the ice particle
rise velocity and the ice concentration source term coefficient $\gamma$
significantly affect the frazil ice dynamics. 
 In the large rise velocity regime, the ice crystals are rapidly pushed
upward and they cannot act as nuclei for ice formation; in the
opposite regime of small rise velocity they tend 
to be transported  downwards in warmer regions where they tend to melt down.
The maximum of ice production is obtained 
in those situations where the rise velocity is 
of the same order of magnitude of the peak rise velocity $\wmax$.
We have developed a simple model which 
captures the trend of the growth rate as a function
of the relevant parameters.
The model has been derived in the approximation
where the ice is produced far beneath the water surface,
however it is found to be valid also for free surface flow.

The parameter 
which plays a key role in fixing the concentration
value at the statistically steady state is the 
heat exchange source term coefficient $\beta$.
The greater
is $\beta$, the lower is the domain averaged ice concentration.
Moreover, the presence of more frazil ice crystals induces a rapid fall
in the kinetic energy and this energy lowering
increases as the saturated ice concentration increases.

The Boussinesq approximation holds only when frazil
ice mass concentration is very small, i.e. of the same order of
the small non-dimensional parameter $\bar{\alpha} = \alpha \Delta
T$. So, the Boussinesq approximation represents well the
initial stage of ice formation, but it breaks down when the ice concentration
becomes significant. Moreover, at significant ice
concentrations, the particles interaction and the effective
viscosity increment
due to the presence of the ice particles should be taken into
account \cite{DeCarolis05}.

Other effects that were not included in the present study and
which we postpone for the future are: the inclusion
of salinity, the occurrence of different classes
of ice crystals with different size, the 
effect of different conditions at the top of the layer,
for example 
a horizontal wind blowing over the water, or the presence of an ice shelf.

\section*{Acknowledgments}

The work of L.P. was funded by the EU Project 'Convection' and,
partially, by PNRA, the Italian Program for Antarctic Res.
The authors would like to thank
Dr. F. Parmiggiani and Prof. P. Wadhams for valuable comments and
stimulating discussions.

\bibliography{frazil}

\end{document}